# A complete overhaul of the EELS and XAS database:

# eelsdb.eu

Philip Ewels[1], Thierry Sikora[2], Virginie Serin[3], Chris P. Ewels[4] and Luc Lajaunie[4, ‡,*]

*1. Department of Biochemistry and Biophysics, Science for Life Laboratory, Stockholm University, Stockholm 106 91, Sweden*

*2. SAVANTIC AB, Rosenlundsgatan 50, 118 63 Stockholm, Sweden*

*3. CEMES, Université de Toulouse, 29 rue Jeanne Marvig BP 94347 31055 Toulouse, France*

*4. Institut des Matériaux Jean Rouxel, (IMN) – Université de Nantes, CNRS, 2 rue de la Houssinière - BP 32229, 44322 Nantes Cedex 3, France*

[‡]. Present address: *Laboratorio de Microscopías Avanzadas, Instituto de Nanociencia de Aragón, Universidad de Zaragoza, 50018 Zaragoza, España.*

[*]. Corresponding author: *Tel: +34 976 762 985, E-mail address: lajaunie@unizar.es*

BRIEF TITLE: The EELS database: eelsdb.eu

**ABSTRACT**

The Electron Energy-Loss Spectroscopy (EELS) and X-ray Absorption Spectroscopy (XAS) database has been completely rewritten, with an improved design, user interface and a number of new tools. The database is accessible at https://eelsdb.eu/ and can now be used without registration. The submission process has been streamlined to encourage spectrum submissions and the new design gives greater emphasis on contributors' original work by highlighting their papers. With numerous new filters and a powerful search function, it is now simple to explore the database of several hundred of EELS and XAS spectra. Interactive plots allow spectra to be overlaid, facilitating online comparison. An application-programming interface has been created, allowing external tools and software to easily access the information held within the database. In addition to the database itself, users can post and manage job adverts and read the latest news and events regarding the EELS and XAS communities. In accordance with the ongoing drive towards open access data increasingly demanded by funding bodies, the database will facilitate open access data sharing of EELS and XAS spectra.

## 1. INTRODUCTION

Electron energy-loss spectroscopy performed in a (Scanning) Transmission Electron Microscope ((S)TEM) is an unrivalled tool providing access to a wealth of structural, chemical and physical information at the nanoscale. Low-loss EELS spectra can be used to determine the dielectric function, the plasmonic properties, the Young's modulus and for strain mapping of both nanostructures and bulk materials ( Oleshko & Howe, 2007; Arenal et al., 2008; Gu et al., 2010; Oleshko, 2012; Lajaunie et al., 2013; Arenal et al., 2014a). Although XAS and core-loss EELS use a different source (x-rays and electrons, respectively) the nature of the chemical information that can be extracted are similar. Their spectral analyses can yield data on elemental quantification, bonding analysis, and local environment. Much other qualitative and quantitative information can also be obtained from the studies of near-edge fine structure, integrated intensity ratio white lines as well as chemical shifts (Egerton, 2011; S. Zhang et al., 2010; Zhang et al., 2011; Zhu et al., 2013; Lajaunie et al., 2015; d'Acapito et al., 2001; Leon et al., 2004; Banerjee et al., 2005). The main advantage of EELS over XAS is its combination with the TEM, which couples the spectral information with nanometer to sub-angstrom spatial resolution (Zhu et al., 2013; Panchakarla et al., 2015). However, interpretation of EELS and XAS spectra is not straightforward. Low-loss EELS spectra contain many excitation processes including volume and surface plasmons, interband transitions and semi-core states transitions which can confound the identification of specific features (Garcia de Abajo, 2010; Lajaunie et al., 2013; Moreau & Boucher, 2012). XAS and core-loss EELS spectra are also not simple - ionization edges can overlap (Hakouk et al., 2013; Panchakarla et al., 2015) and the physical phenomena behind the white line intensity ratio and near-edge fine structures are highly complex (Krüger, 2010; Mizoguchi et al., 2010; Arenal et al., 2014b; Lajaunie et al., 2015;). Different approaches have been proposed to tackle this (ideally a complete study should include multiple approaches (Danet et al., 2010; Arenal et al., 2014b; Lajaunie et al., 2015)). One is the use of *ab-initio* calculations (Lajaunie et al., 2013; Arenal et al., 2014b), however they are time consuming and require prior knowledge of the crystallographic structure. Another is the construction of models stacking all the possible spectral components (Verbeeck & Van Aert, 2004; L. Zhang et al., 2010), but this assumes that all spectral components involved are known and can be reliably modeled (Egerton, 2011).

One of the most common approaches to overcome these difficulties is to use a fingerprinting procedure. This requires a catalogue of reference spectra and can be used for quantification, phase identification and interpretation of spectral features (Garvie et al., 1994; Calvert et al., 2005; Arevalo-Lopez & Alario-Franco, 2009; Danet et al., 2010; Wang et al., 2012). The fingerprinting method is fast and reliable (Danet et al., 2010; Hakouk et al., 2013) although comparison of spectra taken with different experimental parameters should be taken with caution.

To fulfill the requirement for a catalogue of reference spectra, the EELS database was created in the late 1990s (Sikora & Serin, 2008), hosted by the CEMES laboratory. It gathered more than 200 spectra covering 35 elements of the periodic table and rapidly became the largest open-access electronic repository of spectra from Electron Energy-Loss Spectroscopy and X-ray Absorption Spectroscopy (XAS) experiments. The EELS database is now a common tool used by spectroscopists, theoreticians, students from both academic and private organisations (L. Zhang et al., 2010; Shumilova et al., 2011; Núñez-González et al., 2010).

Since the initial development of the EELS database, the utilisation of online data resources has undergone a revolution in many different scientific disciplines, including materials science. Driven by increases in Internet access speeds and improvements in available software tools, web sites such as the Materials Project (Jain et al., 2013) and the Crystallography Open Database (Gražulis et al., 2012) allow interactive searching of large online databases via a range of different materials properties. The incorporation of application-programming interfaces (API) allows a blurring of the line between offline and online software, allowing programs and websites to access each other's databases in order to 'add value' to user activities. In parallel there is an ongoing drive towards Open Science and Open Data, increasingly imposed by funding agencies, ensuring free online access to scientific results and source data. At the same time there are ever increasing demands on the time of researchers, and web tools now need to be fast to learn, intuitive, and easy to access and update.

In this new global context, a complete overhaul of the EELS database was deemed necessary. The database website has been rewritten from scratch and is now accessible at https://eelsdb.eu/. Unlike the previous site, the new EELS database can be browsed without requiring registration. The design and user interface has been greatly improved and a number of new features have been implemented, including interactive plots and an API. This paper will describe these new improvements and features.

## 2. MATERIALS AND METHODS

### 2.1 DATABASE AND WEBSITE CONSTRUCTION

Besides the migration of spectral data and their metadata, the new EELS database is a completely new website. It has been constructed around a custom installation of WordPress (WordPress, 2015) which offers a number of advantages; the core package provides content-management functionality for handling website pages and news updates. Numerous community plugins power some of the simpler features on the site such as job advertisements (WP Job Manager, 2015) and e-mail contact forms (Contact Form 7, 2015). Other plugins try to minimise spam and improve security, whilst the website forum is built using the bbPress plugin (bbPress, 2015). Perhaps the most critical feature of WordPress is its flexibility. A large and powerful code library drives the core of WordPress; custom themes and plugins are able to tie into this code base to drive custom features and functionality. The EELS database uses a custom theme to build the output and aesthetics of the website, whilst a custom plugin creates the pages needed to browse, view and submit spectra to the database. This separation between content and design allows future updates to change independently without affecting each other.

The spectra database plugin was written using the WordPress custom post types API. The submission form is completely bespoke and allows complete control over the interface for website users. There is a second form visible only to website administrators which allows spectra to be edited without the same strict requirements. This approach allows features that would not otherwise be possible, such as integrated registration of new users during the submission process. Spectra browsing and filtering is achieved using a complex set of meta-variable query logic combined with WordPress title and taxonomy searches. Spectra are saved as flat files on the server, with metadata held in the database. Future development with server-side HyperSpy integration (Peña et al., 2015) may enable format homogenisation and improved consolidation between file and database metadata.

The view spectrum page renders the HTML from a template file using the WordPress core, built with PHP and MySQL. Once loaded, bespoke JavaScript code loads the spectrum data from a file on the server. The data is parsed and plotted using HighCharts – an open-source JavaScript graphing library (HighCharts, 2015). Plots generated using HighCharts do not require any special browser plugins and work on virtually all devices.

## 2.2 APPLICATION PROGRAMMING INTERFACE

To facilitate access to the new EELS database by other software applications and websites, an application-programming interface has been created. This works much like the regular website – a URL is requested from the server and a response given. However, instead of the typical response used by web browsers to create a human-friendly web page, the API returns data in a machine-friendly format. The API is accessed via https://api.eelsdb.eu/which lists available API methods. EELS spectra, users and news can be searched and retrieved. Responses give metadata and permalinks, including links to download the spectrum data. Currently, the API does not require any kind of authentication and is read-only; spectra must be submitted manually through the main website, but in future it may be possible to develop routes whereby other software such as EELS acquisition programs are able to submit spectra to the database directly. The API makes it possible for a wide range of tools developed by other groups to use the EELS database, greatly increasing the utility and visibility of spectra contained within.

## 3.   RESULTS AND DISCUSSION

From the homepage, users are guided to the three main sections of the website: news, browse spectra and spectra submission. The main text shows the number of spectra currently available and the number of atomic elements covered by the database. The blue menu at the top of the screen is used to navigate to the other sections of the website (Figure 1). These include a calendar (displaying events related to electron microscopy, EELS and other spectroscopies), a links page, a jobs section, a user forum and a user profile management page. No registration is required to browse the website and database. This will result in better visibility for contributors, as spectra and contributor profiles will be indexed by internet search engines. Users can create a free account to post in the forum, advertise employment opportunities and upload spectra.

### 3.1 UPLOAD

The upload procedure has been completely rewritten in order to simplify the process. The spectral data files can be uploaded in .msa, .csv or .txt format. Gatan Digital Micrograph files (.dm3 and .dm4 mainly) are not yet supported; the authors hope to offer a server-based version of HyperSpy to automatically handle file conversion in the future. In addition to providing the data file, several meta-data fields must be filled during the upload process. These cover sample description (name, elemental composition, purity…), spectra type (low-loss, core-loss, XAS…), description of the microscope and experimental conditions (incident beam energy, collection and convergence angles…) and any data treatments which were performed (corrections for gain and dark current, deconvolution methods…). The fields related to the microscope and experimental conditions can be saved under a named preset (blue rectangle, Figure 2) and recalled in one click during future submissions (red rectangle, Figure 2). Contributors are able to highlight the ionization edges present in their spectra using a selection tool with a periodic table to select the atomic element. Upon clicking an atomic element, appropriate edges can be selected from a list providing edge energies (Figure 3). The level energies are based on those listed in Egerton, 2011. It is possible to associate spectra by specifying their URLs, indicating that they have been taken in the same conditions and on the same sample. This is particularly useful when linking low-loss and core-loss spectra that have been taken on the same area of the sample. Zero-loss spectra can also be uploaded and associated in order to allow other researchers to perform their own data treatments and deconvolutions. Finally, bibliographic references for the publication corresponding to the data can be added. This step has also been simplified: after providing the Digital Object Identifier (DOI) and clicking on the *"Find ref"* button, the website will search crossref (crossref, 2015) and attempt to automatically fill all of the bibliography fields. If a dataset does not have an associated publication, users are invited to upload their data to a service such as Zenodo (Zenodo, 2015). Such providers can assign a DOI to the dataset, which can then be logged within the EELS DB. This makes it easier for downstream users to cite usage of the data. Once the spectrum is submitted, all meta-data fields are visible on the spectrum's page and can be used to filter spectra on the browse page, a clickable link directly to the source publications where available.  It is our expectation that easily available open access spectra directly linked to their source publications will encourage use and citation of data submitted to the site.

**3.2 BROWSE AND DISPLAY**

The browse page allows researchers to view the full list of spectra available in the database (over two hundred spectra at the time of writing). To narrow the search, spectra can be filtered using a range of fields. Clicking the *"Show / Hide Filters"* button at the top right of the page (red rectangle, Figure 4) reveals tools to select the nature of the spectrum (core-loss, low-loss...), the nature of the ionization edge ($K$, $L_{2,3}$, $M_{2,3}$…), the energy range, the energy resolution, the use of a monochromator and more. The exact sample formula can be specified, or constituent atomic elements can be selected using a periodic table. The filters are applied by using the *"Filter Spectra"* button (blue rectangle, Figure 4) and the resulting spectra are shown below. Several attributes are displayed in columns for each spectrum. By default, the table is ordered by specimen formula, but it can be sorted by any other attribute by clicking on the different column titles. A spectrum of interest can be displayed by clicking its name, or a new search can be performed by using the *"Clear Filters"* button.

When viewing a spectrum, the data is plotted in-page using an interactive plotting toolset (Figure 5). Plots can be zoomed by clicking and dragging, or axis limits manually defined using text boxes. All plots have a small menu button in the top right, which gives the option for the graph to be downloaded in a range of publication-ready formats. Registered users are able to save 'favourite' spectra, which are shown in a drop-down box when viewing other spectra. Selecting an item in this list overlays that spectrum in the plot, facilitating online comparison before downloading. Beneath the plot, meta-data specified by the author is shown. This includes links to associated spectra and bibliographic information when available. Finally, each spectrum page provides links to content with a matching atomic formula in the Materials Project maintained by the Lawrence Berkeley National Laboratory (Jain et al., 2013). This yields fast and simple access to a wealth of supplementary information about the material, including crystallographic structure and electronic band structure.

## 3.1 EELS DB QUERIES FROM EXTERNAL TOOLS

The extensive API included with the new EELS database website allows external tools to easily access the information held within the database. Currently, the API provides an interface to the news articles, the list of registered users and the uploaded spectra. Spectra can be queried using the full range of filters available through the normal web page. By using these access points, tools written by third parties can directly incorporate EELS database spectra into their code. This is beneficial to all parties - users of these tools have easier access to the spectra to help with their research, whilst contributors to the EELS database have increased visibility for their work.

As an example of how the API can be used, we see the recent integration into HyperSpy: an open-source python library commonly used for multi-dimensional data processing and visualisation of EELS spectra. After initial development by both parties, EELS database integration will soon be available as part of the core HyperSpy package. In practice, this allows HyperSpy users to retrieve and plot spectra of the EELS database from within their analysis script. The code is fast, simple and will travel with the script, making it an ideal approach for reproducible research. In its simplest form, usage comes down to a single line of code. For example, to plot all core loss spectra held in the EELS database which contain Iron and Oxygen, a HyperSpy user can include the following line in their script:

```
utils.plot.plot_spectra(datasets.tem_eels.eelsdb(element=["Fe", "O"],
type="coreloss"))
```

## 3.4 OPEN SCIENCE AND RESEARCH FUNDING

The term 'Open Science' refers to a drive towards greater access to public research data (Open Data) and publications (Open Access). Open Science policy is promoted by many international and national organizations (OECD, 2015). As a direct consequence, several research funding agencies such as the Welcome Trust, the National Science Foundation and H2020 programs require (or strongly encourage) that data generated by funded research should be freely available to the public.

The structure and data contained within the EELS database are licensed under an Open Data Commons Open Database (ODb) license. During submission, contributors agree to license their data under the ODb license. It is the responsibility of contributors to ensure that they own the intellectual property for the raw data that they upload. Generally publishers retain copyright over manuscripts, but not raw data. However the original data may be subject to rights claimed by third parties (if the research is funded by a company for instance) (Carroll, 2015). The ODb license has been specifically designed to protect data and databases. It states that anyone is free to copy, distribute, transmit and adapt the data, as long as the original work of the contributors and the EELS database are credited. If the data are altered or built upon, the result can only be distributed under the same license. The rest of the website is licensed under the Creative Commons Attribution Share Alike 2.0 license (CC BY-SA). These licenses ensure that the EELS database is fully compliant to the principles of Open Science. As a result of this, the EELS database has been recognized as an Open Data research repository by the global Registry of Research Data Repositories organization (Pampel et al., 2012). The EELS database can thus be included in data management plans, fulfilling an increasingly common requirement found within numerous funding agency grant proposal processes.

## 4. CONCLUSIONS

Here, we describe the creation of a new iteration of the EELS database, constructed around a custom installation of WordPress and accessible at https://eelsdb.eu/.

The spectral upload procedure has been greatly simplified, with appropriate meta-data making spectra much easier to cite. Data can be viewed in-page using an interactive plotting toolset which allows zooming and overlay of multiple spectra, which facilitates online comparison before downloading. We note that although the focus of this current article has been on EELS data, the database is also home to XAS spectra which can be analysed, viewed, uploaded and treated using the same tools as described above.

In addition, an extensive API has been included which allows external tools to easily access the information held within the database. Currently the API is used by the HyperSpy software allowing HyperSpy users to retrieve and plot EELS database spectra from within their analysis script. We hope that other tools will also use this API now that it is public, and that future collaborations can lead to ever tighter integration into existing tools.

Finally the database structure and data contained within it are licensed under an Open Data Commons Open Database license. This ensures that the EELS database is fully compliant to the principles of Open Science and can be included in data management plans required by funding agencies.

**ACKNOWLEDGEMENTS**

The authors acknowledge the following entities for the funding: the *Institut des Matériaux Jean Rouxel* (IMN, Nantes, France) and the *Centre d'Elaboration de Matériaux et d'Etudes Structurales* (CEMES, Toulouse, France) laboratories, the European microscopy network *ESTEEM 2*, the French microscopy network *METSA* and the French microscopy society *Sfµ*. The authors warmly acknowledge everyone who has contributed to the database and the beta-testers as well who greatly helped us to improve the website. L.L would like to thank Dr. P. Moreau (IMN) for the opportunity to work on this project.

**FIGURE LEGENDS**

Figure 1. *(Color online, two-columns width)* Homepage of the EELS and XAS database: https://eelsdb.eu.

Figure 2. *(Color online)* Section of the spectrum submission page. All fields related to the microscope and experimental conditions can be saved under a named preset (blue rectangle) and recalled in one click during future submissions (red rectangle).

Figure 3. *(Color online, two-columns width)* Ionization edge selection tool used to select edges visible in the uploaded spectra. The atomic element is selected using a periodic table and edges are picked from the resulting list.

Figure 4. *(Color online, two-columns width)* The browse spectra page. Filters are shown on the browse spectra page by clicking the *"Show / Hide filters"* (red rectangle) and then activated by using the *"Filter Spectra"* button (blue rectangle). The table of results is displayed below.

Figure 5. *(Color online, two-columns width)* The display page of a XAS spectrum of diamond (Jaouen et al., 1995). The interactive plot is shown, with an overlaid EELS spectrum of "glassy" graphite EELS (submitted by Pr. D. Muller, University of Cornell).

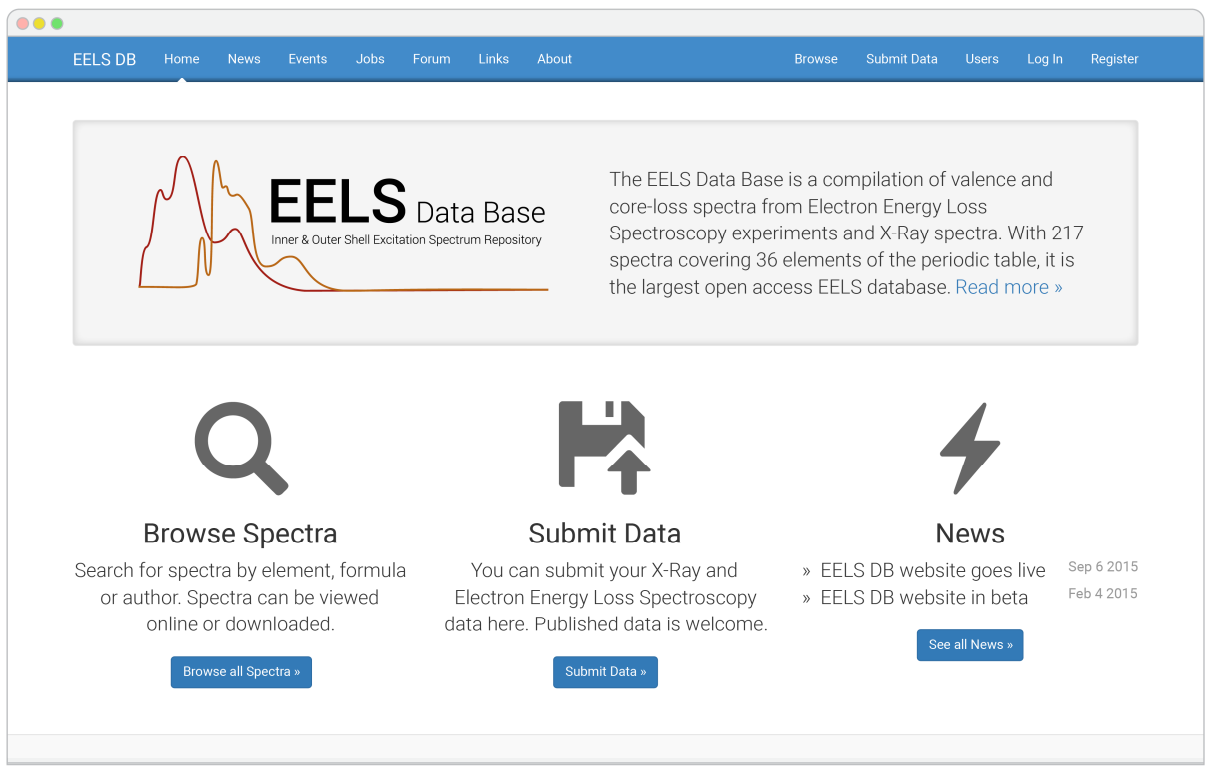

Figure 1

Figure 2

Figure 3

Figure 4

# Diamond Formula: C  `X Ray Abs`

View in Materials Explorer    Download

Submitted by Michel Jaouen, April 20, 2005.

**Source / Purity:** CVD grown Diamond - Purity : 100% C

Author Comments: Pre-edge substracted Analyst: M. Jaouen, G. Tourillon. Temperature: 300. BEAM/Crystal: none.

Share Axis    Remove overlay plot

### EELS Spectra
Click and drag in the plot area to zoom in

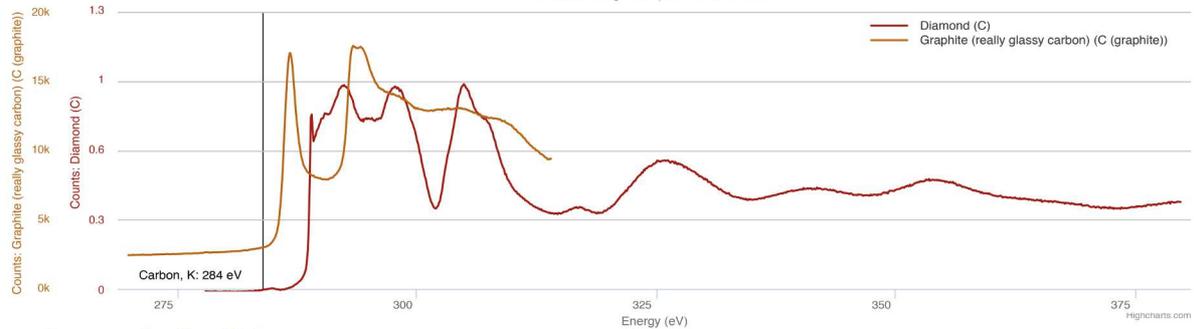

You can manually set the axis limits.

## Spectrum Metadata

| | | | | | |
|---|---|---|---|---|---|
| Specimen Name | Super-ACO (LURE) | Microscope Name / Mo... | Super-ACO (LURE) | Dark Current Correction | No |
| Spectrum Type | `X Ray Abs` | Gun Type | ABS3 (SA22) | Reference | M. Jaouen, G. Tourillon, J. |
| Specimen Formula | C | Incident Beam Energy | 800 kV | | Delafond, N. Junqua, G. Hug, |
| Data Range | 277.8 eV - 379.69 eV | Dispersion | 0.1 eV/pixel | | Diamond & Related Materials |
| Source and Purity | CVD grown Diamond - Purity: | Acquisition Mode | XAS Electron yield | | Vol.4 200-206 (1995) |
| | 100% C | Beam Current | 180 | | |
| Elemental Edges | C,K | Detector | TGM (1200 lines mm-1 | | |
| Alternative URL | Old EELS DB | | grating) | | |

Download Spectrum

Figure 5